\documentclass{emulateapj}
\usepackage{apjfonts}

\newcommand{\pivec}{\mbox{\boldmath $\pi$}}

\lefthead{JUNG ET AL.}
\righthead{REANALYSIS OF MACHO 97-BLG-41}

\begin{document}
\title{Reanalysis of the Gravitational Microlensing Event MACHO-97-BLG-41 based on Combined Data}

\author{
Youn Kil Jung$^{1}$,
Cheongho Han$^{1,}\footnotemark[4]$,
Andrew Gould$^{2}$,
and 
Dan Maoz$^{3}$
}

\footnotetext[4]{Corresponding author}

\bigskip\bigskip
\affil{$^{1}$Department of Physics, Institute for Astrophysics, 
  Chungbuk National University, Cheongju 371-763, Korea}
\affil{$^{2}$Department of Astronomy, Ohio State University, 
  140 West 18th Avenue, Columbus, OH 43210, USA}
\affil{$^{3}$School of Physics and
  Astronomy, Tel-Aviv University, Tel Aviv 69978, Israel}

\begin{abstract}
MACHO-97-BLG-41 is a gravitational microlensing event produced by 
a lens composed of multiple masses detected by the first-generation 
lensing experiment.  For the event, there exist two different 
interpretations of the lens from independent analyses based on two 
different data sets: one interpreted the event as produced by 
a circumbinary planetary system while the other explained the light 
curve with only a binary system by introducing orbital motion of the lens. 
According to the former interpretation, 
the lens would be not only the first planet detected via microlensing 
but also the first circumbinary planet ever detected. 
To resolve the issue using 
state-of-the-art analysis methods, we reanalyze the event based on 
the combined data used separately by the previous analyses.
By considering various higher-order effects, we find that the 
orbiting binary-lens model provides a better fit than the circumbinary 
planet model with $\Delta\chi^2\sim 166$. The result signifies the 
importance of even and dense coverage of lensing light curves in the 
interpretation of events.
\end{abstract}


\section{INTRODUCTION}

The last two decades have witnessed tremendous progress in gravitational
microlensing experiments. On the observational side, improvement in both
hardware and software has contributed to the great increase of the
detection rate of lensing events from tens of events per year at the
early stage of the experiments to thousands per year at the current
stage. In addition, photometry based on
difference imaging substantially improved the quality of photometry.

Along with the observational progress, there also has been advance on the
analysis side. A good example is the analysis of light curves of lensing
events produced by multiple masses. The light curve of a single-lens
event is described by a simple analytic equation with a small number of
parameters and the lensing magnification varies smoothly with respect to the
lensing parameters. As a result, observed light curves can be easily
modeled by a simple $\chi^2$ minimization method. However, when 
events are produced by multiple masses, modeling light curves becomes very
complex not only because of the increased number of lensing parameters
but also because of the non-linear variation of lensing magnification
with respect to the parameters. The non-linearity of lensing
magnifications is caused by the formation of caustics which denote
positions on the source plane where the lensing magnification of a point
source diverges. Caustics cause difficulties in lens
modeling in two ways. First, they make it difficult to use a simple
linearized $\chi^2$ minimization method in modeling light curves because of 
the complexity of the parameter space caused by the singularity.  Second,
magnification computations for source positions on a caustic are
numerically intensive. As a result, modeling a multiple
lens event was a daunting task at the early stage of lensing experiments. 
However, with the introduction of efficient non-linear modeling methods 
such as the Markov Chain Monte Carlo (MCMC) algorithm, combined with 
advances in computer technology such as computer 
clusters or graphic processing units, precise and fast modeling became possible and now it 
is routine to model light curves in real time as lensing events progress 
\citep{dong06, cassan08, kains09, bennett10, ryu10, bozza12}. 
Furthermore, current modeling take into account various subtle higher-order effects.

In this paper, we reanalyze the lensing event MACHO-97-BLG-41 that is a
multiple-lens event detected by the first-generation lensing experiment 
Massive Compact Halo Objects \citep[MACHO:][]{alcock93}.  For the
event, there exist two different interpretations. Based mainly on the
data obtained from the MACHO experiment, \citet{bennett99} interpreted 
the event as produced by a circumbinary planetary system where a 
planet was orbiting a stellar binary. On the other hand, \citet{albrow00}, 
from independent analysis based on a different data set obtained 
by the Probing Lensing Anomalies NETwork \citep[PLANET:][]{albrow98} 
group, arrived at a different interpretation that the light curve 
could be explained without the introduction of a planet but rather by 
considering the orbital motion of the binary lens. According to the 
interpretation of \citet{bennett99}, the lowest-mass component of the 
triple-mass lens would be not only the first planet detected via microlensing 
but also the first circumbinary planet ever detected. 
Despite the importance of the event, the issue of its correct 
interpretation remains unresolved. 
Therefore, we revisit MACHO-97-BLG-41, 
applying state-of-the-art analysis methods to the combined data 
used separately by \citet{bennett99} and \citet{albrow00}.

\begin{deluxetable}{lrlr}
\tablecolumns{9}
\tablewidth{0pc}
\tablecaption{Data sets\label{table:one}}
\tablehead{
\multicolumn{2}{c}{MACHO data} & 
\multicolumn{2}{c}{PLANET data} \\
\colhead{observatory} & 
\colhead{number} & 
\colhead{observatory} & 
\colhead{number}
} 
\startdata
MSO 1.3 m  ($R$) & 711 & SAAO 1.0 m      ($I$) & 97 \\
MSO 1.3 m  ($B$) & 772 & SAAO 1.0 m      ($V$) & 14 \\
MSO 1.9 m  ($R$) & 16  & ESO/Dutch 0.9 m ($I$) & 58 \\
Wise 1.0 m ($R$) & 17  & ESO/Dutch 0.9 m ($V$) & 18 \\
CTIO 0.9 m ($R$) & 35  & Canopus 1.0 m   ($I$) & 95 \\
                 &     & Canopus 1.0 m   ($V$) & 14 \\
                 &     & CTIO 0.9 m      ($I$) & 49 \\
                 &     & Perth 0.6 m     ($I$) & 26 
\enddata
\tablecomments{MSO: Mount Stromlo Observatory; 
CTIO: Cerro Tololo Inter-American Observatory, 
SAAO: South Africa Astronomy Astronomical Observatory, 
ESO: European Southern Observatory.}
\end{deluxetable}

\begin{figure*}[ht]
\epsscale{0.8}
\plotone{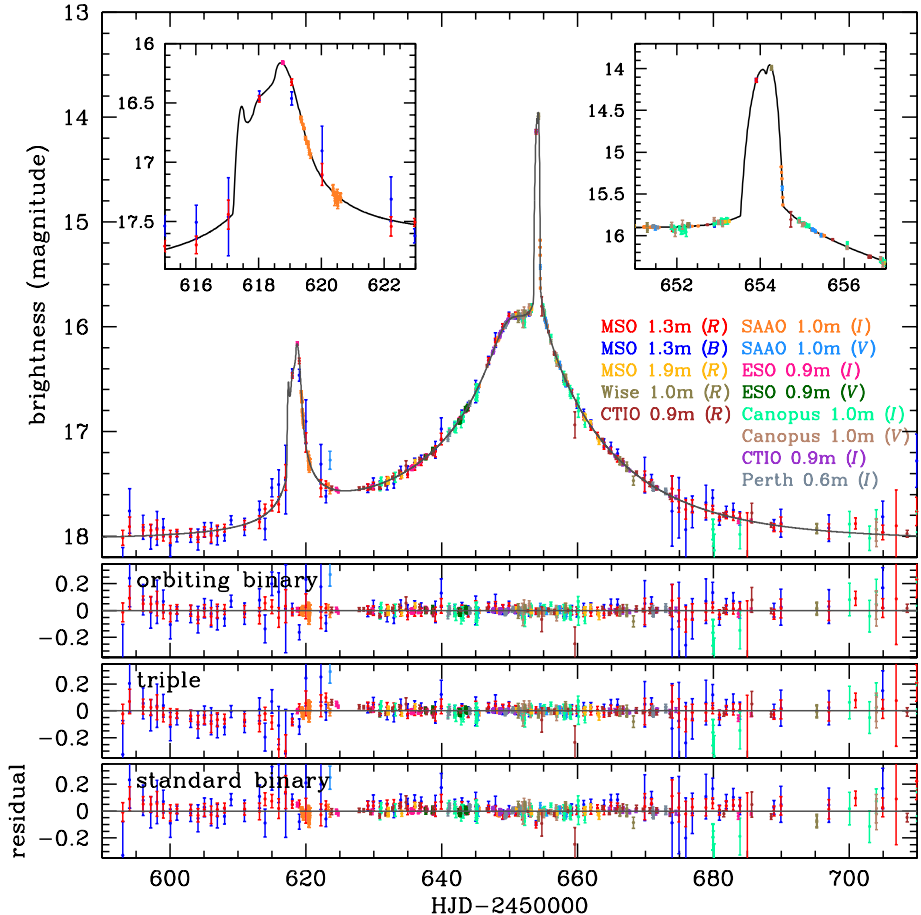}
\caption{\label{fig:one}
Light curve of MACHO-97-BLG-41 based on the combined MACHO plus 
PLANET data sets.  Also presented is the best-fit model curve from 
our analysis.  The two insets in the upper panel show the enlarged 
view of the two caustic-involved features at ${\rm HJD'} \sim 619$ 
and 654.  The lower three panels show the residual from the orbiting 
binary, triple, and static binary lens models.
}\end{figure*}

\section{Data}

The data used for our analysis come broadly from two streams. The first
stream comes from MACHO survey observations plus the Global Microlensing
Alert Network \citep[GMAN:][]{alcock97} and Microlensing Planet Search
\citep[MPS:][]{rhie99} follow-up observations. We refer to this data
set as the ``MACHO data''. The other stream comes from observations
conducted by the PLANET group. We denote this data set as the ``PLANET 
data''.  In Table~\ref{table:one}, we present telescopes, passbands, 
and the number of data of the individual data sets. Note that 
\citet{bennett99} analyzed only the MACHO data set while \citet{albrow00} 
conducted their analysis based only on the PLANET data set.

In order to use data sets obtained from different observatories, we
rescale error bars. For this, we first readjust error bars so that the
cumulative distribution of $\chi^2$/dof ordered by magnifications matches 
to a standard cumulative distribution of Gaussian errors by introducing 
a quadratic error term \citep{bachelet12}. Then, error bars are rescaled 
so that $\chi^2$/dof becomes unity, where $\chi^2$ is derived from the 
best-fit solution. During this normalization process, $3\sigma$ outliers 
from the solution are removed to minimize their effect on modeling.

In Figure~\ref{fig:one}, we present the light curve 
based on the combined MACHO + PLANET data. The light curve 
is characterized by two separate peaks at ${\rm HJD}'= {\rm HJD}-2450000 
\sim 619$ and 654. We note that the models of \citet{bennett99} and 
\citet{albrow00} are consistent in the interpretation of the overall 
light curve including the peak at ${\rm HJD}' \sim 654$. However, the 
two models differ in the interpretation of the peak at ${\rm HJD}'
\sim 619$. While \citet{bennett99} explained the first peak by introducing 
an additional lens component of a circumbinary planet, \citet{albrow00} 
described the peak by considering the orbital motion of the binary lens.

\begin{deluxetable*}{l|rrr|rr}
\tablewidth{0pt}
\tablecaption{Best-fit Parameters\label{table:two}}
\tablehead{
\multicolumn{1}{c|}{parameter} &
\multicolumn{3}{c|}{binary lens} & 
\multicolumn{2}{c}{triple lens} \\
\multicolumn{1}{c|}{} &
\multicolumn{1}{c}{standard} &
\multicolumn{1}{c}{orbit} &
\multicolumn{1}{c|}{orbit+parallax} &
\multicolumn{1}{c}{standard} &
\multicolumn{1}{c}{parallax}
}
\startdata
$\chi^2$/dof                     &  2355.8/1920       & 1915.1/1920        &  1912.4/1920       &  2086.6/1920       & 2077.3/1920         \\
$t_0$ (HJD$'$)               &  653.519$\pm$0.004 &  653.426$\pm$0.006 &  653.426$\pm$0.007 &  653.371$\pm$0.008 &   653.388$\pm$0.008 \\
$u_0$ $(10^{-2})$                &  8.123$\pm$0.024   &  -7.346$\pm$0.079  &  7.303$\pm$0.074   &  7.407$\pm$0.047   &   7.284$\pm$0.049   \\
$t_{\rm E}$ (days)               &  20.26$\pm$0.016   &   24.11$\pm$0.229  &  23.95$\pm$0.213   &  24.82$\pm$0.155   &   24.50$\pm$0.150   \\
$s_1$                            &  0.494$\pm$0.0003  &  0.479$\pm$0.002   &  0.481$\pm$0.002   &  0.485$\pm$0.002   &   0.480$\pm$0.002   \\
$q_1$                            &  0.481$\pm$0.003   &   0.346$\pm$0.005  &  0.342$\pm$0.005   &  0.318$\pm$0.005   &   0.326$\pm$0.005   \\
$\alpha$                         &  1.211$\pm$0.003   &  -1.176$\pm$0.003  &  1.179$\pm$0.003   &  1.152$\pm$0.003   &   1.156$\pm$0.005   \\
$s_2$                            &                    &                    &                    &  1.900$\pm$0.007   &   1.875$\pm$0.008   \\
$q_2$ $(10^{-3})$                &                    &                    &                    &  5.149$\pm$0.160   &   4.644$\pm$0.180   \\
$\psi$                           &                    &                    &                    &  1.872$\pm$0.003   &   1.858$\pm$0.008   \\
$\rho_*$ $(10^{-3})$             &  9.69$\pm$0.08     &  7.49$\pm$0.12     &  7.91$\pm$0.11     &  7.36$\pm$0.09     &   7.35$\pm$0.01     \\
$\pi_{{\rm E},N}$                &                    &                    &  -1.59$\pm$0.38    &                    &  0.21$\pm$0.15      \\
$\pi_{{\rm E},E}$                &                    &                    &   0.29$\pm$0.10    &                    &  0.34$\pm$0.08      \\
$ds/dt$ $({\rm yr}^{-1})$        &                    &  -0.72$\pm$0.050   &  -0.67$\pm$0.05    &                    &                     \\
$d\alpha/dt$ $({\rm yr}^{-1})$   &                    &  -0.95$\pm$0.037   &  -0.30$\pm$0.16    &                    &                
\enddata
\end{deluxetable*}

\section{Analysis}

\subsection{Standard Binary-lens Model}

To describe the observed light curve of the event, we begin with a
standard binary-lens model. Basic description of a binary-lens event
requires seven lensing parameters. The first three of these parameters describe
the geometry of the lens-source approach including the time of the
closest lens-source approach, $t_0$, 
the lens-source separation (normalized by the angular Einstein
radius of the lens, $\theta_{\rm E}$) at that 
moment, $u_0$, and the time scale for the source to cross the Einstein 
radius, $t_{\rm E}$ (Einstein time scale). Another three parameters describe 
the binary nature of the lens including the projected separation $s$ (in 
units of $\theta_{\rm E}$) and the mass ratio $q$ between the binary-lens 
components, and the source trajectory angle with respect to the binary 
axis $\alpha$. The two peaks of the light curve of MACHO-97-BLG-41 are 
likely to be features involved with caustic crossings or approaches 
during which finite-source effect becomes important 
\citep{dominik95,gaudi99,gaudi02}. To account for this effect, 
an additional parameter, the normalized source radius 
$\rho_*=\theta_*/\theta_{\rm E}$ is needed, where $\theta_*$ is the 
angular source radius. In our standard binary-lens modeling, we
additionally consider the limb-darkening variation of the source star
surface brightness by introducing linear-limb-darkening 
coefficients, $\Gamma_\lambda$. The surface brightness profile 
is modeled as $S_\lambda \propto¡ð1-\Gamma_\lambda(1-3\cos\phi/2)$,
where $\lambda$ denotes the observed passband and $\phi$ is the 
angle between the normal to the surface of the source and the line 
of sight toward the source \citep{albrow99}. Based on the source type (subgiant) 
determined by the spectroscopic observation conducted by \citet{lennon97}, 
we adopt coefficients from \citet{claret00}. The adopted values are 
$\Gamma_B=0.793$, $\Gamma_V=0.666$, $\Gamma_R=0.575$, and $\Gamma_I=0.479$ 
for data sets acquired with a standard filter system. For the MSO 1.3m data, which 
used a non-standard filter system, we adopt $(\Gamma_B+\Gamma_V)/2$ for 
the $B$-band data and $(\Gamma_R+\Gamma_I)/2$ for the $R$-band data.

Although the two sets of solutions presented by \citet{bennett99} and 
\citet{albrow00} already exist, we separately search for solutions in 
the vast parameter space encompassing wide ranges of binary separations 
and mass ratios in order to check the possible existence of other solutions. 
From this, we find a unique solution with $s\sim 0.49$ and $q\sim 0.48$. 
See Table \ref{table:one} for the complete solution of the model. 
We note that the solution is basically consistent with the results of \citet{bennett99} 
and \citet{albrow00} in the interpretation of the main part 
of the light curve including the peak at ${\rm HJD}' \sim 654$. At the bottom panel
of Figure \ref{fig:one}, we present the residual of the standard binary model. It is
found that there exist some significant residuals for the standard model. This is
also consistent with the previous analyses that a basic binary model is
not adequate to precisely describe the light curve.

\begin{figure*}[ht]
\epsscale{0.8}
\plotone{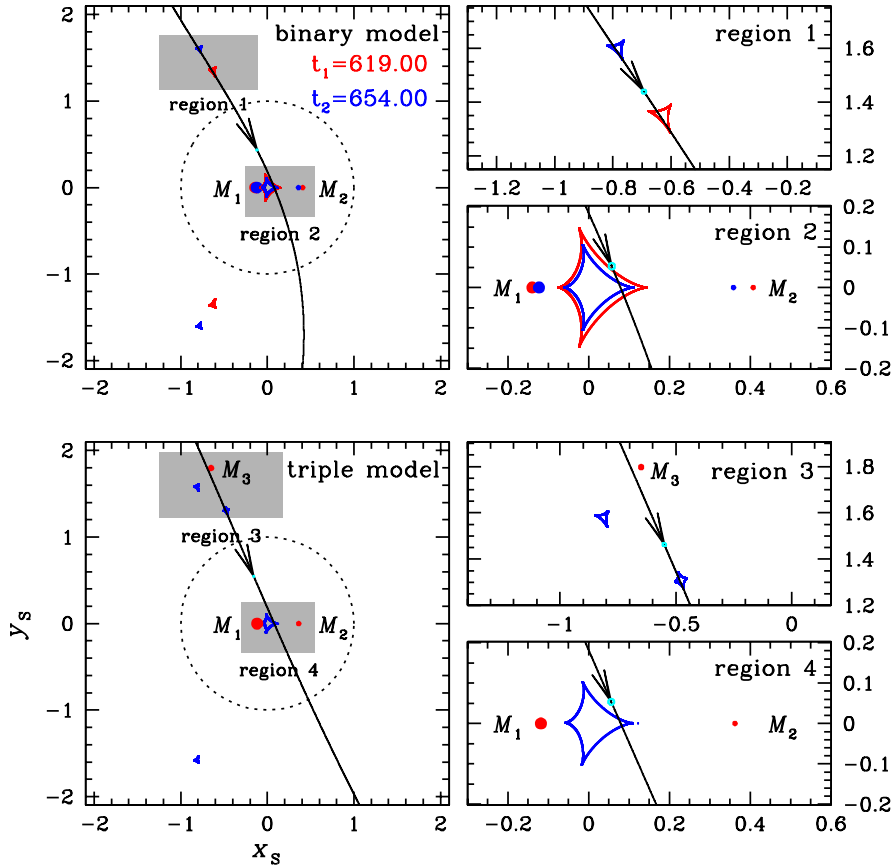}
\caption{\label{fig:two}
Geometry of the lens system for the best-fit binary (upper panels) and 
triple (lower panels) lens solutions.  In each panel, the closed cuspy 
figures represent caustics and the curve with an arrow is the source 
trajectory.  The filled circles is the locations of the lens components 
and the dotted circle is the Einstein ring centered at the barycenter 
of the lens.  Right panels show the enlarged view of the corresponding 
shaded regions in the left panels. For the binary model, caustics and 
lens positions vary in time due to the orbital motion.  We present two 
sets of caustics at ${\rm HJD}'=619$ and 654.  The coordinates are co-rotating 
with the binary axis so that the binary axis aligns with the abscissa. 
All lengths are scaled by the Einstein radius $\theta_{\rm E}$.
}
\end{figure*}

\subsection{Higher-order Effects}

The existence of residuals in the fit of the standard binary lens model 
suggests the need for considering higher-order effects. We consider 
the following effects.

First, we consider the effect of the motion of an observer caused by 
the orbital motion of the Earth around the Sun. This ``parallax'' 
effect causes the source trajectory to deviate from rectilinear, 
resulting in long-term deviations in lensing light curves  
\citep{gould92}. The event MACHO 97-BLG-41 lasted $\sim$ 100 days, which 
is an important portion of the Earth's orbital period, i.e. 1 year, 
and thus the parallax effect can be important. Considering the parallax 
effect requires two additional parameters $\pi_{{\rm E},N}$ and 
$\pi_{{\rm E},E}$, that are the two components of the lens parallax 
vector $\pivec_{\rm E}$ projected onto the sky along the north and east 
equatorial coordinates, respectively. The magnitude of the parallax 
vector corresponds to the relative lens-source parallax, 
$\pi_{\rm rel}={\rm AU}(D_{\rm L}^{-1}-D_{\rm S}^{-1})$, scaled to the 
Einstein radius of the lens, i.e. $\pi_{\rm E}=\pi_{\rm rel}/\theta_{\rm E}$ \citep{gould04}.

Second, the orbital motion of a binary lens can also cause the source 
trajectory to deviate from rectilinear \citep{dominik98}. The orbital motion causes 
further deviations in lensing light curves by deforming the caustic over 
the course of the event. The ``lens orbital'' effect can be important 
for long time-scale events produced by close binary-lens events for which 
the event duration comprises an important portion of the orbital period 
of the lens system \citep{shin13}. To first order approximation, the 
lens orbital motion is described by two parameters, $ds/dt$ and $d\alpha/dt$, 
that represent the change rates of the normalized binary separation and the 
source trajectory angle, respectively \citep{albrow00}.

Third, we also check the possible existence of a third body in the lens 
system. Introducing an additional lens component requires three additional 
lensing parameters including the normalized projected separation, $s_2$, 
and the mass ratio, $q_2$, between the primary and the third body, and the 
position angle of the third body with respect to the line connecting the 
primary and secondary of the lens, $\psi$.

\subsection{Result}

We test models considering various combinations of the higher-order effects. 
In Table~\ref{table:two}, we list the goodness of the fits and the best-fit 
parameters for the individual tested models. From the comparison of the models, 
we find the following results.

First, we confirm the result of \citet{albrow00} that the consideration of the 
orbital effect substantially improves the fit. We find that the improvement 
is $\Delta\chi^2\sim 441$ compared to the static binary model.  When we 
additionally consider the parallax effect, the improvement 
of the fit, $\Delta\chi^2\sim 2.7$, is very meager. This implies that among 
the two effects, the orbital motion of the lens is the dominant higher-order 
effect in explaining the residual from the standard model. In 
Figure~\ref{table:one}, we present the model light curve on the top of the 
observed light curve and the residual of the model. In Figure~\ref{fig:two}, 
we also present the geometry of the lens system.

Second, we find that the existence of a third body does not provide a fully 
acceptable fit. Our best-fit solution of three-body lens modeling is consistent 
with the solution of \citet{bennett99} in the sense that the third body is a 
circumbinary planet with a small mass ratio. 
See Table~\ref{table:two} for the best-fit parameters and 
Figure~\ref{fig:two} for the lens system geometry.  With the introduction of a 
planetary third body, the fit does improve from the standard binary model with 
$\Delta\chi^2\sim 270$. The additional consideration of the parallax effect 
further improves the fit with $\Delta\chi^2\sim 9$, but 
it is still substantially poorer than the 
orbiting binary-lens model with $\Delta\chi^2\sim 166$. From the comparison 
of the residual (see Figure~\ref{fig:one}), it is found that the triple lens 
model cannot precisely describe the light curve during and before the first 
peak, $595 \lesssim {\rm HJD'}\lesssim 625$.

The key PLANET data that exclude the triple lens are from SAAO.
Their smooth ``parabolic'' decline signals a caustic exit along
the axis of cusp. This is compatible with the triangular
caustic generated by the close binary, but not with the
quadrilateral caustic induced by the putative planet. In particular,
the near-symmetric shape of the quadrilateral caustic implies
that a cusp-axis trajectory would have a near-symmetric excess flux
in the approach to the first peak as after its exit, which is
not seen in the pre-peak MSO data.

\section{Conclusion}

We have conducted a reanalysis of the event MACHO-97-BLG-41 for which there exist two 
different interpretations. From the analysis considering various higher-order 
effects based on the combined data sets used separately by the previous analyses, 
we find that the dominant effect for the deviation from the standard binary-lens 
model is the orbital motion of the binary lens.

The result signifies the importance of even and dense coverage of lensing light 
curves for correct interpretation of gravitational lenses. For MACHO-97-BLG-41, 
the difference between the two previous interpretations partially stems from 
the poor coverage of the first peak that is important in the interpretation.  
Although a strategy based on survey plus follow-up observations can densely cover 
anomalies occurring at an expected time (e.g., peak of a high-magnification event) 
or long-lasting anomalies, it would be difficult to densely cover short-lasting 
anomalies arising abruptly at an unexpected moment.
Since the first-generation lensing experiments, there has been great progress 
in survey experiments. The cadence of survey observations has increased from 
$\sim$1 -- 2 per day to several dozens a day for the current lensing experiments 
(OGLE: \citet{udalski03}, MOA: \citet{bond01}, \citet{sumi03}, Wise: \citet{shvartzvald12}). 
Furthermore, a new survey based on a network of multiple telescopes (KMTNet: Korea 
Microlensing Telescope Network) equipped with large format cameras 
is planned to achieve a cadence of more than 100 per day. 
With improved coverage, the characterization of microlenses by 
future surveys will be more accurate.

\acknowledgments
Work by CH was supported by Creative Research Initiative Program
(2009-0081561) of National Research Foundation of Korea.
AG was supported by NSF grant AST 1103471. DM and AG acknowledge 
support by the US Israel Binational Science Foundation.

\end{document}